# Motor neuron pathology in CANVAS due to *RFC1* expansions

Vincent Huin, M.D., Ph.D.[1,2,†], Giulia Coarelli, M.D.[1,3,†], Clément Guemy, M.D.[1], Susana Boluda, M. D., Ph.D.[1,4], Rabab Debs, M.D.[5], Fanny Mochel, M.D., Ph.D.[1,3], Tanya Stojkovic, M.D.[6], David Grabli, M.D., Ph.D.[5], Thierry Maisonobe, M.D.[6], Bertrand Gaymard, M.D., Ph.D.[7], Timothée Lenglet, M.D.,[7] Céline Tard, M.D.[2,8], Jean-Baptiste Davion, M.D.[2,8], Bernard Sablonnière, M.D., Ph.D.[2], Marie-Lorraine Monin, M.D.[1], Claire Ewenczyk, M.D. Ph.D.[1,3], Karine Viala, M.D.[6], Perrine Charles, M.D., PhD[1,3], Isabelle Le Ber, M.D., Ph.D.[1,9], Mary M Reilly, M.D., Ph.D.[10], Henry Houlden, M.D., Ph.D.[10], Andrea Cortese, M.D., Ph.D.[10], Danielle Seilhean, M.D. Ph.D,[1,4], Alexis Brice, M.D.[1], and Alexandra Durr, M.D., Ph.D.[1,3]

## Abstract

CANVAS caused by *RFC1* biallelic expansions is a major cause of inherited sensory neuronopathy. Detection of *RFC1* expansion is challenging and CANVAS can be associated with atypical features.

We clinically and genetically characterized 50 patients, selected based on the presence of sensory neuronopathy confirmed by EMG. We screened *RFC1* expansion by PCR, repeat-primed PCR, and Southern blotting of long-range PCR products, a newly developed method. Neuropathological characterization was performed on the brain and spinal cord of one patient.

Most patients (88%) carried a biallelic (AAGGG)$_n$ expansion in *RFC1*. In addition to the core CANVAS phenotype (sensory neuronopathy, cerebellar syndrome, and vestibular impairment), we observed chronic cough (97%), oculomotor signs (85%), motor neuron involvement (55%), dysautonomia (50%), and parkinsonism (10%). Motor neuron involvement was found for 24 of 38 patients (63.1%). First motor neuron signs, such as brisk reflexes, extensor plantar responses, and/or spasticity, were present in 29% of patients, second motor neuron signs, such as fasciculations, wasting, weakness, or a neurogenic pattern on EMG in 18%, and both in 16%. Mixed motor and sensory neuronopathy was observed in 19% of patients. Among six non-RFC1 patients, one carried a heterozygous AAGGG expansion and a pathogenic variant in *GRM1*. Neuropathological examination of one RFC1 patient with an enriched phenotype, including parkinsonism, dysautonomia, and cognitive decline, showed posterior column and lumbar posterior root atrophy. Degeneration of the vestibulospinal and spinocerebellar tracts was mild. We observed marked astrocytic gliosis and axonal swelling of the synapse between first and second motor neurons in the anterior horn at the lumbar level. The cerebellum showed mild depletion of Purkinje cells, with empty baskets, torpedoes, and astrogliosis characterized by a disorganization of the Bergmann's radial glia. We found neuronal loss in the vagal nucleus. The pars compacta of the *substantia nigra* was depleted, with widespread Lewy bodies in the *locus coeruleus*, *substantia nigra*, hippocampus, entorhinal cortex, and amygdala.

We propose new guidelines for the screening of *RFC1* expansion, considering different expansion motifs. Here, we developed a new method to more easily detect pathogenic *RFC1* expansions. We report frequent motor neuron involvement and different neuronopathy subtypes. Parkinsonism was more prevalent in this cohort than in the general population, 10% *versus* the expected 1% ($p<0.001$). We describe, for the first time, the spinal cord pathology in CANVAS, showing the alteration of posterior columns and roots, astrocytic gliosis and axonal swelling, suggesting motor neuron synaptic dysfunction.

**Author affiliations:**

1 Sorbonne Université, Paris Brain Institute, APHP, INSERM, CNRS, Paris, France

2 Univ. Lille, Inserm, CHU Lille, U1172 - LilNCog (JPARC) - Lille Neuroscience & Cognition, F-59000 Lille, France

3 AP-HP, Pitié Salpêtrière University Hospital, Genetics Department, Sorbonne University, Paris, France

4 Laboratoire Neuropathologie Raymond Escourolle, AP-HP, Pitié Salpêtrière University Hospital, Sorbonne University, Paris, France

5 AP-HP, Pitié Salpêtrière University Hospital, Department of Neurology, Sorbonne University, Paris, France

6 Institut de Myologie, Centre de Référence de Pathologie Neuromusculaire Paris-Est, AP-HP, Pitié Salpêtrière University Hospital, Sorbonne University, Paris, France 7 AP-HP, Pitié Salpêtrière University Hospital, Department of Neurophysiology, Sorbonne University, Paris, France

8 Centre de Référence des Maladies Neuromusculaires, CHU Lille, F-59000 Lille, France

9 AP-HP, National Reference Center for "Rare and Young Dementia", IM2A, Pitié-Salpêtrière University Hospital, Sorbonne University, Paris, France

10 Department of Neuromuscular Disease, UCL Queen Square Institute of Neurology and The National Hospital for Neurology and Neurosurgery, London, UK

[†] These authors contributed equally to this work as first co-authors.

Correspondence to: Pr Alexandra Durr

Paris Brain Institute, Pitié-Salpêtrière Paris CS21414, 75646 PARIS Cedex 13, France

alexandra.durr@icm-institute.org

tel. +33 1 57 27 46 82, fax +33 1 57 27 47 26

**Running title**: New pathological features of CANVAS

# Introduction

Sensory neuronopathies (or ganglionopathies) are a subgroup of sensory neuropathies caused by the degeneration of neurons in the dorsal root ganglia, followed by secondary degeneration of ascending fibers of the posterior column in the spinal cord. The diagnosis of ganglionopathy is based on nerve conduction studies that show severe impairment of sensory conduction in the upper and lower limbs or its absence. The most frequently inherited forms are Friedreich ataxia (OMIM #229300), caused by intronic GAA biallelic expansions in *FXN*; sensory ataxic neuropathy, ptosis, and ophthalmoparesis (OMIM #607459), caused by *POLG* biallelic pathogenic variants; and CANVAS syndrome, caused by an intronic AAGGG biallelic expansion in *RFC1*.[1] CANVAS is an adult-onset, slowly progressive disorder that shows autosomal recessive transmission.[2] Series of CANVAS patients of different ethnicity have been reported[3–6], describing the phenotype. The evidence of novel repeat motifs, $(ACAGG)_{exp}$[7], $(AAGAG)_{exp}$[7], and $(AGAGG)_{exp}$[6], suggests a dynamic nature of the pentanucleotide expansion in the *RFC1* gene.

The core phenotype associated with CANVAS includes sensory neuronopathy (constant), cerebellar ataxia, and a reduced visually enhanced vestibulo-ocular reflex. Additional clinical features include chronic cough, usually arising years before the onset of other symptoms, abnormal eye movements, and cerebellar, cerebral, and spinal cord atrophy by MRI.[4]

Detection of the biallelic expansion in CANVAS is technically challenging because of (i) the relatively high frequency of non-pathogenic expansions of $(AAAAG)_{exp}$ and $(AAAGG)_{exp}$, and probably other motifs, with a cumulative allelic frequency of approximately 24% in the general population,[1] and (ii) the large size of the pathogenic expansions, requiring Southern blotting of genomic DNA to detect the two pathogenic alleles. The Southern-blot method is time-consuming, requires a significant amount of input DNA ($\geq 5$ μg), and is highly dependent on DNA quality.

We aimed to clinically and genetically characterize 50 patients that we selected based on the presence of sensory neuronopathy, confirmed by nerve conduction studies, and for whom Friedreich ataxia and *POLG*-related disorders were excluded. We optimized the detection of pathogenic expansions in the *RFC1* gene and performed exome sequencing to identify other genetic causes of ganglionopathy.

# Materials and methods

## Patients

We recruited 50 patients from 37 families with sensory neuronopathy from a historical case database of ataxic patients without molecular diagnosis through the two National Reference Centers for Rare Diseases in Paris, neurogenetic and neuromuscular ($n = 42$), and in Lille, neuromuscular ($n = 8$). Index cases were selected on the presence of a sensory neuronopathy based on an electrophysiological examination and a Camdessanché score $> 6.5$.[8] All patients had abolished or severely decreased sensory nerve action potential (SNAPs) in four limbs. Then, we included apparently affected siblings too. All patients were Caucasian, except for two from North Africa.

Patients were examined by a neurologist or geneticist with expertise in neurogenetics and/or movement disorders (G.C, R.D, F.M, T.S, D.G, T.M, B.G, C.T, JB.D, ML.M, C.E, K.V, P.C, I.LB, A.B, and A.D) and went through a diagnostic work up, including cerebral and spine MRI (to exclude common differential diagnoses, i.e. cerebral small vessel disease, radiculopathy, compressive myelopathy) and the exclusion of known causes of neuronopathy, such as vitamin B12 or E deficiency, toxic medications, and immune conditions. Dysimmune neuropathy was excluded when the clinical history was indolent or when other necessary workup was negative (CSF analysis and plexus MRI).

These patients without causal diagnosis were highly suspected to suffer from a sensory neuronopathy caused by a genetic disease. GAA expansions in the *FXN* gene were ruled out for all patients using short-florescent PCR and triplet-primed PCR. Screening of *POLG* was performed for those with cerebellar syndrome using Sanger or next-generation sequencing (panel for 359 ataxia genes, Supplementary Methods). We have performed an extensive molecular screening according to the clinical features and/or familial history including repeat expansion in SCA genes, mitochondrial DNA analyses and different targeted sequencing using next generation sequencing and/or exome sequencing.

Cerebellar syndrome was scored by the assessment and rating of ataxia (SARA, max. value: 40)[9] and disability stage, measured using the SPATAX disability score (0: no functional handicap, 1: no functional handicap but signs at examination, 2: mild, able to run, unlimited walking, 3: moderate, unable to run, limited walking without aid, 4: severe, walking with one stick, 5: walking with two sticks, 6: unable to walk, requiring a wheelchair, 7:

confined to a bed). Vestibular function was assessed by oculomotor recordings or otological examination, including a video head-impulse-test and the search for vestibular areflexia.

The pallesthesia at ankles was examined by quantitative evaluation using a Rydel-Seiffer tuning fork (C64) with a scale from 0/8 (worse) to 8/8 (best). We evaluated the deep tendon reflexes by the National Institute of Neurological Disorders and Stroke (NINDS) grading scale: 0 (absent reflex), 1 (small reflex, less than normal, or obtained with reinforcement), 2 (lower half of normal reflex), 3 (upper half of normal reflex), and 4 (pathological increased reflex).[10] We defined the upper motor neuron involvement as the presence of brisk and diffused reflexes at upper and lower limbs and/or extensor plantar reflex and/or spasticity. We defined the the lower motor neuron involvement as the presence of diffused fasciculations, and/or diffused wasting, and/or diffuse proximo-distal chronic neurogenic pattern (characterized by large and high motor unit potential amplitudes) on EMG with or without rest activity.

## Oculomotor recording

Participants were seated in a dark, silent room, with their head held in place by chin-level and forehead supports. Horizontal and vertical movements of both eyes were recorded with a video-based eye tracker (Eyebrain eye tracker) at a sampling rate of 300 Hz, allowing the accurate analysis of saccades. Paradigms consisted of a visually guided saccade task (reflexive saccades triggered toward a suddenly presented visual target), an anti-saccade task (saccade performed in the opposite direction of a visual target), and a horizontal smooth pursuit task, in which the subject was asked to accurately follow a target moving with a sine wave motion at 12 and 25°/s maximum velocities. We recorded saccade latency, amplitude, and velocity, the accuracy of both centrifugal and centripetal saccades, the percentage of errors (*i.e.*, saccades initially triggered towards the target) in the anti-saccade task, pursuit abnormalities, the presence and type of nystagmus, and the presence of square wave jerks. A visually enhanced vestibulo-ocular reflex (VVOR) was sought only for the group of patients for whom vestibular impairment was suspected.

## *RFC1* analysis

The identification of biallelic AAGGG expansions in *RFC1* was performed as presented in the flowchart (Fig. 1).

We performed PCR with fluorescent primers as described previously[1], with the following exceptions: the forward primer was fluorescently-labelled with 6-carboxyfluorescein (6-FAM) and the PCR premix included 10% DMSO. After PCR amplification, the length of the amplified PCR products was analyzed by capillary electrophoresis on an ABI3730 DNA analyzer (Applied Biosystems®, Saint-Aubin, France). We considered an allele to be normal if the PCR product consisted of a single peak without PCR slippage, a fluorescence intensity > 500 arbitrary units, and a length of 348 bp (*i.e.* reference allele $(AAAAG)_{11}$) ± 10 bp, which corresponds to normal alleles with 9 to 13 pentanucleotide (AAAAG) repeat units.

In the absence of amplification, suggesting the presence of two expanded alleles, we performed repeat-primed PCR (RP-PCR), as previously described[1], to detect the presence of at least one pathogenic $(AAGGG)_n$ motif. Then, we confirmed the presence of a homozygous expanded pathogenic allele $(AAGGG)_n$ by Southern blotting of long-range PCR (LR-PCR) products, similarly to other methods commonly used for the molecular diagnosis of other repeat-expansion diseases.[11] We adapted the LR-PCR reported by Cortese *et al.*[1] adding agarose gel electrophoresis of LR-PCR products, capillary transfer to a nylon membrane, and Southern blotting of long-range PCR-products (Fig. 2).

## Southern blotting of long-range PCR products

This new method uses the Expand™ Long Template PCR System for long amplifications (Roche Diagnostics, Meylan, France). Each 50 µl reaction consisted of 1X Expand long PCR buffer n°1, 0.35 mM dATP, dCTP, and dTTP, 0.23 mM dGTP, 0.12 mM 7-deaza-GTP, 0.25 µM Long-Range flanking PCR forward 3'-TCAAGTGATACTCCAGCTACACCGTTGC-5' and reverse 3'-GTGGGAGACAGGCCAATCACTTCAG-5' primers, 3.75 U Taq DNA polymerase Expand Long, and either 20 or 4 ng total genomic DNA. Amplifications were performed in a MyCycler™ (Bio-Rad, Marnes-la-Coquette, France) (Supplemental Table S1). PCR products were resolved on 1% agarose gels stained with ethidium bromide. Three DNA molecular weight standards were loaded on each agarose gel: DNA markers IV, V, and VII (digoxin-labeled) (Roche Diagnostics, Meylan, France). Gel electrophoresis lasted 16 h with the power at 40 V, followed by Southern blot transfer and probe-mediated detection. Briefly, agarose gels were denatured in 0.4 M NaOH, 0.6 M NaCl, and rinsed twice in 2X SSC before being transferred to Hybond N+ nylon membranes. Blots were prehybridized in 5X SSC, 5% liquid block solution, and 0.02% SDS for 30 min at 59°C. Hybridization was performed in a fresh solution containing 40 pmol $(CCCTT)_5$ probe end-labeled with digoxin for 2 h at 59°C, followed by two washes in 5X SSC, 0.1% SDS at 42°C for 5 min, and two washes in 1X SSC, 0.1% SDS. Detection was performed using ECL™ reagents and Hyperfilm-ECL according to the manufacturer's instructions (Amersham-Pharmacia Biotech, Orsay, France). Film exposure times ranged from 5 to 30 min. Image acquisition during the control of electrophoresis and film fluorescent exposure using Epi illumination was performed with an ImageQuant™ LAS 4000 (GE Healthcare, Buc, France).

On the control of electrophoresis image, normal alleles could be detected (≈348 bp), as well as expanded alleles with low somatic instability, corresponding either to polymorphic expansions with $(AAAAG)_n$ or $(AAAGG)_n$ motifs. On the Southern blots (Fig. 2), expanded *RFC1* alleles appeared in most cases as smears or multiple fragments, corresponding to probable somatic heterogeneity and/or contraction of repeated regions during PCR cycling. Polymorphic expansions were not detected until a film exposure time of 45 min. A faint band around 348 bp, corresponding to the normal alleles, could be detected in healthy and heterozygous subjects. Such band may probably be caused by a cross hybridization: a complementary base pairing between the $(CCCTT)_5$ probe and the non-expansed alleles $(AAAAG)_n$ although these sequences are not perfectly identical. The PCR products of healthy subjects were free of smears on the blot.

## Exome sequencing

We performed exome sequencing for the six patients without *RFC1* biallelic expansions (Supplementary methods).

## Neuropathology

A post-mortem examination was performed on patient KEN-334-6, who signed an informed consent form, for the French National Brain Bank Network Donation Program NeuroCEB. The postmortem delay was 11 h. The right half of the brain, including the hemisphere, brainstem, and cerebellum, was fixed by immersion in 4% formaldehyde (10% formalin). The contralateral hemisphere was frozen at −80°C. Formalin-fixed brain samples from selected regions, including the cerebral cortex, hippocampus, basal ganglia, cerebellum, brainstem, spinal cord, and anterior and posterior roots of the spinal nerves, were embedded in paraffin and cut at a thickness of 3 μm. The spinal cord was sampled at the cervical, thoracic, and lumbar levels. The sections were deparaffinized in graded alcohol solutions and stained with hematoxylin-eosin (HE) and HE combined with Luxol fast blue for myelin. Selected sections were immunostained using a Ventana BenchMark stainer (Roche™, Tucson, AZ, USA). The biotinylated secondary antibody was included in the detection kit (Ventana Medical Systems Basic DAB Detection Kit 250-001). Diaminobenzidine (DAB) was used as a chromogen. The pretreatment and antibodies used for immunohistochemistry are listed in Supplemental Table S2. Genomic DNA was isolated from cerebellum, according to standard procedures.

## Data availability

The authors confirm that the data supporting the findings of this study are available within the article and its supplementary materials.

## Results

Eighty-eight percent (44/50) of our patients from 32/37 families with sensory neuronopathy carried a biallelic pathological (AAGGG)$_n$ expansion in the *RFC1* gene. Nineteen came from a family compatible with recessive transmission of the disease, 10 from five families with apparent dominant transmission, and 15 were isolated cases. One patient was heterozygous for a pathological (AAGGG)$_n$ expansion and five had no expansion in *RFC1*. Variants in non-*RFC1* expanded patients are provided in supplementary data and supplemental Tables S3-S4.

### *RFC1* biallelic expansion carriers (Table 1)

The carriers of biallelic expansions were 22 men and 22 women. Detailed clinical data were available for 38 patients, with a mean age at onset of 53.3 ± 10.6 years and a mean disease duration of 14.6 ± 10.7 years. The mean disability stage was 3.3 ± 1.3 out of a maximum of 7 (bedridden) and mean the SARA score 12.7 ± 8.6 ($n = 28$) out of a maximum of 40. The core CANVAS features (sensory neuronopathy, cerebellar syndrome and vestibular impairment) were present for 13 of the 38 patients, whereas association of both sensory neuronopathy and cerebellar syndrome were present for 37 of the 38 patients. The neuronopathy was characterized by: i) a severely decreased sense of vibration at the ankles (1-3/8 by the quantitative evaluation using a Rydel-Seiffer tuning fork) in six cases (16.2%) and completely abolished (0/8) in the remaining 31 (81.5%), ii) pin-prick hypoesthesia for 17 patients (44.7%) and neuropathic pain for 20 (52.6%), iii) decreased (1/4 NINDS score) or abolished (0/4 NINDS score) reflexes at the ankles for 28 patients (73.6%), the patella for 10 (26.3%), and the upper limbs for seven (18.4%); iv) *pes cavus* for seven patients (18.4%), and scoliosis for two (64 and 56 years old). Nerve conduction studies were performed, showing a dramatic decrease or the absence of sensory action potentials for all patients except for one case (AAD-1016-2), who had only a decreased vibration sense at the ankles and oculomotor abnormalities evocative of cerebellar dysfunction. Cerebellar syndrome was characterized by gait impairment, dysarthria in 20 patients (52.6%), and limb dysmetria in eight (21%). We found vestibular involvement in 14/16 (87.5%) of cases: seven patients had an altered VVOR (four have been recorded), six had an abnormal otological examination, and one had both. In addition, a dry spasmodic chronic cough was reported for 33/34 patients for which this data was available (97.1%). Most of the time, it preceded the onset of other neurological signs by decades. Dysautonomia was present for 19 patients (50%), with mainly bladder dysfunction ($n = 13$, 34.2%), but other signs were also reported, such as erectile dysfunction ($n = 5$), orthostatic hypotension ($n = 5$), Raynaud phenomenon ($n = 3$), xerostomia ($n = 2$), hypersalivation ($n = 2$), xerophthalmia ($n = 1$), profuse sweating ($n = 1$), and hemicrania paroxystica ($n = 1$).

Interestingly, motor neuron involvement was found for 24 of 38 (63.1%) patients for whom the information was available. Both types of motor neurons were affected in six (15.7%), first motor neurons alone for 11 (28.9%) (brisk reflexes scored 4/4 at NINDS scale, extensor plantar reflexes and/or spasticity) and second motor neurons for seven (18.4%). Mixed motor and sensory neuronopathy was observed for seven patients (18.9%), with moderate decreases of motor action potential and/or electromyography evocative of denervation. Only four patients presented either with spasticity ($n = 3$) or motor deficit ($n = 1$).

Four (10.5%) had parkinsonism (bradykinesia in combination with rigidity or rest tremor), with a mean age at examination of 74.2 years. Dopamine transporter (DaT) SPECT was reduced for all four. L-Dopa treatment slightly improved the symptoms of three of four patients. We found a higher prevalence of parkinsonism relative to that of the general population[12] (10% *versus* 1%, $p < 0.001$, Chi-square test). Dystonia was reported for only one patient. Moderate cognitive impairment was present for four patients (10.5%), with a Montreal Cognitive Assessment score between 13 and 23 (maximum value 30). Hearing loss was reported for six patients (15.7%). Additional signs included ptosis, reported for three patients (7.9%), and epilepsy for two (generalized tonic-clonic seizures for AAD-1016-2 and simple partial seizures for AAR-727, both appeared in childhood).

## Oculomotor examination

Oculomotor examination data were available for 34/44 RFC1 carriers. Oculomotor signs of cerebellar involvement were frequent and clinically reported in 29/34 RFC1 patients (85.3%). Nystagmus was found in 27 cases (79.4%): downbeat ($n = 13$), gaze-evoked ($n = 11$), or combined ($n = 3$). Saccadic pursuit was noted in 22 patients (64.7%), saccade dysmetria in 15 (44.1%), fixation instability in nine (26.4%), and slow saccades in eight (23.5%). Ten patients underwent oculomotor recording. The saccades were dysmetric in all patients except one and latency and velocity were impaired in six patients. Nystagmus was constant: gaze-evoked in all and downbeat in half. Anti-saccades errors were also frequent (8/10). Visually enhanced vestibulo-ocular reflex, when tested during the oculomotor recording, was always impaired ($n = 4$).

## Brain MRI

Cerebral MRI was available for 30 patients. Cerebellar atrophy was reported in 22 cases (73.3%), involving the vermis in all and the hemispheres in 14. Diffuse cortical and subcortical atrophy was also reported in five cases, possibly linked to age (mean age at examination 74 ± 6.6 years). No other radiological abnormalities were reported in CANVAS patients, except for isolated hyperintensity of the right middle cerebellar peduncle in one case and possible pons atrophy in another.

## Neuropathological findings

We performed a post-mortem neuropathological examination of RFC1 patient KEN-334-6, who died of respiratory distress due to bilateral pneumonia at 74 years of age. His parents were not related. His father died at the age of 80 and suffered from Parkinson's disease and his mother died at the age of 100, without neurological disease. One of his three sisters had cramps and sensory neuropathy. She died at 70 years of age from neoplasia. His first symptom was cramps at 25 years of age. At 45, he developed mixed sensory and cerebellar ataxia, as well as vestibular areflexia and neuropathic pain. Sensory neuronopathy was confirmed by nerve conduction study. He had brisk reflexes, extensor plantar reflex, diffused fasciculations, and cramps in the lower limbs. Parkinsonism, rapid eye movement sleep behavior disorder, dysautonomia (bladder dysfunction, erectile dysfunction, and orthostatic hypotension), and chronic cough enriched the phenotype, appearing progressively from the age of 65. His parkinsonism was improved by L-Dopa treatment and he developed motor fluctuations and dyskinesia six years later. He started to show cognitive impairment at around the age of 70 (his MOCA score was 10/30 at 73) and to have visual hallucinations at 73. A DaTscan showed dopaminergic denervation and brain MRI cortical and subcortical atrophy at 73.

At the postmortem examination, his brain weighted 1,400 grams. The hemispheres were of normal size, whereas there was atrophy of the cerebellar vermis that affected the lingula, culmen, and central lobe. Moderate dilation of the lateral ventricle was observed, predominantly on the occipital horn. There was no atrophy of the hippocampus, amygdala, or basal ganglia. Examination of the brainstem showed pallor of the *substantia nigra*. The *locus coeruleus* was not identifiable. The inferior olive was of normal size. Cross sections of the spinal cord showed pallor and atrophy of the posterior columns that were more pronounced at the cervical level. SMI-310 phosphorylated neurofilament immunostaining showed axonal loss that was more prominent in the fasciculus gracilis than the fasciculus cuneatus (Fig. 3A-3B) and milder pallor of the vestibulospinal and spinocerebellar tracts (Fig. 3A). The corticospinal tracts appeared normal. Despite the clinical signs of upper and lower motor neurons involvement, we did not observe neuronal loss from the anterior or lateral horns, but axonal swelling was observed at the contact with motor neuron bodies (Fig. 3C). TDP43 and p62 immunohistochemistry did not show skein-like inclusions in motor neurons in the anterior horns. MBP/2F11 double immunolabeling showed atrophic posterior lumbar roots and normal anterior lumbar roots (Fig. 3D-E).

In the cerebellar hemispheres, there was mild depletion of Purkinje cells, with a number of empty baskets and torpedos and a certain amount of Bergmann layer gliosis, with no signs of demyelination (Fig. 3G). The loss of Purkinje cells and the presence of empty baskets, torpedos, and astrogliosis were more severe in the vermis. The cerebellar dentate nucleus was normal, as were the inferior olivary nuclei. Pontine nuclei showed no neuronal loss but neuronal depletion was observed in the vagal nuclei (especially dorsal motor nucleus of the vagus and the solitary nucleus). CD68 and CD163 antibodies showed foci of microglial activation in the vestibular and vagal nuclei (Fig. 3F). Neuronal depletion was obvious in the pars compacta of the *substantia nigra*. α–synuclein immunoreactive Lewy bodies and Lewy neurites were found in the medulla oblongata (dorsal motor nucleus of the vagus, solitary nucleus, reticular formation)*, locus coeruleus*, and *substantia nigra* (Fig. 3H). Abundant α-synuclein immunoreactivity was observed in the hippocampus, amygdala, entorhinal (Fig. 3I), and anterior cingular cortices, α-synuclein aggregates were sparse in the neocortex including the frontal and parietal cortices. These lesions were consistent with a neocortical Lewy pathology as per the latest neuropathological diagnostic criteria[13] and consistent with the neuropathological diagnosis of PD. Tau inclusions were detected in the entorhinal cortex and hippocampus, and were probably age-related (Braak stage 2). The subthalamic nucleus, striatum, pallidum, and thalamus were normal. The frontal cortex showed vacuolization of the external layers with mild neuronal loss. A few superficial neurons showed translocation of TDP-43 from the nucleus to the cytoplasm. The calcarine sulcus was unremarkable. β-amyloid immunohistochemistry (IHC) was negative in all regions tested.

The severity of the neurological disease contrasted with the few neuronal loss, that is why we performed further immunostainings to study the astrocytes in the patient (Fig. 4A-E,K,N), a healthy control (Fig. 4F-J,M,P) and a ALS patient (Fig. 4L,O). The age at post-mortem examination were 74, 71, and 95 in KEN-334-6 patient, ALS patient, and healthy control respectively. Immunohistochemical study by double labeling of GFAP (Fig. 4A) and non-phosphorylated neurofilaments (SMI-32) (Fig. 4B) showed marked astrocytic gliosis, with protoplasmic cell bodies and numerous positive extensions in contact with dendrites. In the cerebellar, the astrogliosis was characterized by a disorganization of the Bergmann's radial glia that took on a granular appearance, with, as in the anterior horn, numerous GFAP+ extensions in contact with the dendrites (Fig. 4C-D). Immunohistochemistry of phosphorylated neurofilaments (SMI31) showed disorganization of the parallel fiber network (Fig. 4E). GFAP/AQP4 double labeling showed AQP4 accumulation in astrocytic extensions in contact with motor neurons (Fig 4K), as observed in one ALS patient with similar age at examination (Fig. 4L) and healthy control (Fig. 4M). In the molecular layer of the cerebellum, AQP4 immunostaining involved numerous blistered astrocytic processes (Fig. 4N), compared to ALS patient (Fig. 4O) and healthy control (Fig. 4P).

*RFC1* analysis in the blood and cerebellar samples show highly similar profile between the two tissues with slightly smaller expansion in the cerebellum compared to the blood (≈750 to 790 repeats in the blood *versus* ≈740 to 750 repeats in the cerebellum) (Supplemental Figure S1).

## Discussion

*RFC1* expansions are very large and are comprised of various motifs, making them difficult to detect for diagnostic purposes. Here, we propose new guidelines for the screening of *RFC1* expansions, including the search for all described pathogenic motifs if there is a discrepancy between abnormal fluorescent PCR and normal RP-PCR (Fig. 1). The flow chart takes into account all other recently reported motifs: $(ACAGG)_{exp}$[7], $(AAGAG)_{exp}$[7], $(AGAGG)_n$[6], and $(AAAGG)_{10-25}(AAGGG)_{exp}$.[14] These guidelines should be adapted if other pathogenic motifs or compound heterozygous patients with two different pathogenic motifs are reported in the future.

Given the heterogeneity of the CANVAS locus in *RFC1* intron 1, we also developed a new method, LR-PCR with Southern-blot revelation, to more easily detect the various pathogenic expansions for molecular diagnosis. Most repeat expansions in human diseases can have rare sequence interruptions, such as, for example, in Huntington disease[15], myotonic dystrophy type I[16], and CANVAS.[1] However, such sequence interruptions can lead to difficulties or the failure to detect expansions by TP-PCR or RP-PCR and, thus, the risk of false-negatives. The molecular diagnosis of CANVAS relies strongly on Southern-blotting, which is a time-consuming method and difficult to apply for most diagnostic laboratories for current analyses. Our new methods can detect pathogenic expansion (AAGGG)n and evidenced normal allele as polymorphic expansion. It could be easily adapted to detect other pathogenic motifs using different probes. Lastly, this method does not need another DNA sampling. We propose that LR-PCR with Southern-blot revelation could be easier to use for diagnostic laboratories, as CANVAS ataxia appears to be nearly as common as Friedreich ataxia.

We report a cohort of patients with sensory neuronopathy without causal diagnosis and found biallelic *RFC1* expansions to be the most frequent diagnosis. There is a selection bias in our cohort as our patient were clinically characterized and followed since years and went to extensive molecular screening. The high rate of patients with biallelic *RFC1* expansion does not represent the proportion of CANVAS in sensory neuropathy or ataxic patients. It is not surprising to find *RFC1* as the most frequent diagnosis of sensory neuronopathy in our cohort, given the high frequency of heterozygotes for the expanded $(AAGGG)_{exp}$ allele in the healthy population (0.7%)[1], similar to the frequency of the intronic GAA expanded allele in *FXN*. The rate of the heterozygous expanded $(AAGGG)_{exp}$ allele in our cohort was 2% (1/50 patients). Akçimen *et al*, report a frequency of ≈4% in 163 healthy controls, whereas 15.8% patients from three ataxia cohorts were heterozygous for the pathogenic expansion $(AAGGG)_{exp}$.[6]

In addition to the usual CANVAS phenotype, we also identified a frequent feature, consisting of motor neuron involvement, present in 63% of our RFC1 patients, which was not previously identified[4,5,14,17], despite the fact that motor neuropathy was recently reported for 18/45 (40%) patients[3] and for 13/34 (38%) patients in another cohort.[18] Differently from these two last case series[3,18] that reported a more frequent rate of motor neuropathy than our cohort (18%), we pointed out the presence of fasciculations, wasting, and myokimia. Until now, only one case report of three patients described the presence of fasciculations.[7] In the cohort of 43 RFC1 patients,[17] none presented a sensory-motor neuropathy but despite the sensory neuropathy the deep tendon reflexes were normal (~70% at upper limbs and ~ 45% at knee) or even brisk (~10%). Three of our RFC1 patients were addressed to our neuromuscular reference center for fasciculations as the first signs. In CANVAS, motor conduction is frequently normal, but a moderate decrease in motor action potentials can be observed with preserved muscle strength.[3,19] Prominent first motor neuron involvement is reminiscent of what is described in Friedreich ataxia[20], the major differential diagnosis of CANVAS syndrome, except for the age at onset (Table 2). Friedreich ataxia generally occurs before the age of 25, but late onset Friedreich ataxia, even after the age of 40, can be encountered.[21,22] Among those with late onset Friedreich ataxia, deep tendon reflexes are conserved in 80% of cases and the extensor plantar reflex is present in 55%.[22] The corticospinal tract presents more severe microstructure damage, revealed by diffusion tensor images, in late onset Friedreich ataxia than in the classical form of the disease.[23] The conservation of deep tendon reflexes and H reflexes in CANVAS has been hypothesized to be due to the selective impairment of small and large sensory fibers whereas the muscle spindle fibers (Aα and Aβ) are preservated.[4] However, the presence of pyramidal signs, and not merely the observation of normal deep tendon reflexes, suggests involvement of the pyramidal system, in addition to sensory involvement, as seen in Friedreich ataxia.[21] Clinical differences between late onset Friedreich ataxia and CANVAS are the presence of cough, vestibular impairment, and mild to moderate cerebellar atrophy in CANVAS (Table 2). However, the absence of cerebellar atrophy in CANVAS should not exclude the detection of *RFC1* expansions, similar to Friedreich ataxia.[24]

The neuropathology in Friedreich ataxia primarily affects the large-diameter sensory neurons of the dorsal root ganglia and peripheral sensory nerves, as described for CANVAS.[25] We report the loss of Purkinje cells in this study and it was reported in the original description[25], but they are spared in Friedreich ataxia.[26] We report the first complete brain and spinal cord neuropathological examination of an RFC1 patient with an enriched phenotype, including parkinsonism, dysautonomia, cognitive decline, and visual hallucinations. The examination revealed the prominent involvement of posterior columns and posterior roots of the spinal nerves, as well as milder alteration of the vestibulospinal and spinocerebellar tracts, responsible for the sensory ataxia. The cerebellum showed mild depletion of Purkinje cells. Neuronal loss was found in the vagal nucleus, especially for dorsal motor nucleus and the solitary nucleus. The presence of Lewy bodies and neuronal loss in the *substantia nigra* is consistent with the patient's parkinsonism.

Although the second motor neurons in the anterior and lateral horns were preserved, we found evidence of synaptic dysfunction between the first and second motor neurons. The presence of protoplasmic astrocytes in contact with the cell body of motor neurons, the fragmented aspect of astrocytic processes particularly visible in the molecular layer of the cerebellum, the close contact of astrocytic extensions with the dendrites of motor neurons and Purkinje cells suggest clasmatodendrosis described by Cajal.[27] Although this type of anomaly should be interpreted with caution[28] and requires further investigations, it has recently been proposed that impair synaptic transmission by contact with dendritic spines[29] could offer a plausible explanation for the severity of the neurological disease contrasting with the absence of significant lesions, except those related to Parkinson's disease.

We observed numerous atypical features in our RFC1 patients (Table 1), suggesting that the *RFC1*-associated phenotype may be more complex. Parkinsonism was noted in 10% of patients, confirming the prevalence observed in another RFC1 cohort.[26] This value greatly exceeds the prevalence of Parkinson's disease, the most frequent cause of parkinsonian syndromes. Indeed, in the general population at age 75, this prevalence is ≈1%[12], whereas it was 10% of our patients, whose mean age at examination was 74 ± 6.6 years ($p < 0.001$). All patients with parkinsonism showed presynaptic dopaminergic denervation and, in the single examined case, we found neuronal loss in the *substantia nigra,* Lewy bodies in the *locus coeruleus* and *substantia nigra,* and widespread α-synuclein immunoreactivity. Our results are in line with previous reports[30,31] supporting the parkinsonism as a possible clinical feature in RFC1 patients.

We report detailed oculomotor recording data. Oculomotor abnormalities were present for almost 90% of patients. Recorded patients showed broken pursuit, square wave jerks, dysmetric saccades, and for half, impaired velocity and/or latency. Nystagmus, either gaze-evoked or downbeat, was generally observed. Other oculomotor cerebellar features not yet reported in CANVAS patients[19] were alternating skew deviation and exophoria. Interestingly, almost all patients showed increased anti-saccade task errors, possibly reflecting the cerebellar involvement in executive functions.[32] Vestibular impairment was only investigated when strongly suspected, confirming the abnormal visually enhanced vestibulo-ocular reflex.

Oculomotor abnormalities could evoke *POLG*-related disorders as a differential diagnosis.[33] These patients are younger than CANVAS patients (31 *versus* 54 years at onset) and involvement of the first motor neuron is less frequently reported (11%).[34] Although the mechanism by which *RFC1* expansions cause CANVAS is still unknown, the fact that the two most frequent causes of ganglionopathy in adults (Friedreich ataxia and *POLG* mutations) are both mitochondrial disorders is intriguing. Other inherited causes of sensory neuronopathy are rarer and show phenotypic differences. They include vitamin E-related disorders (AVED and abetalipoproteinemia), spinocerebellar ataxia type 3 (SCA3), *RNF170*-related syndrome[35], mitochondriopathy as myoclonic epilepsy associated with ragged red fibers (MERRF), transthyretin-related familial amyloid polyneuropathy, and diverse rare hereditary neuropathies.[36]

This study confirms *RFC1* as a major cause of sensory neuronopathy and expands the clinical signs associated with CANVAS by the frequent presence of first and second motor neuron involvement. The spinal cord pathology showed the alteration of posterior columns and roots, as well as the marked astrocytic gliosis suggesting motor neuron synaptic dysfunction.

# Acknowledgements

We are grateful to all the patients and family members for their participation in this study. We thank Dr Sylvie Forlani and Ludmila Jornéa (Sorbonne Université, Paris Brain Institute, DNA/Cell Bank), Amélie Labudeck, and Christiane Marzys (UF de Neurobiologie, Lille University Hospital) for their technical assistance.

# Funding

The research leading to these results received funding from the VERUM foundation and "Investissements d'avenir" ANR-11-INBS-0011 – NeurATRIS: Translational Research Infrastructure for Biotherapies in Neurosciences. Andrea Cortese thanks the Medical Research Council (MR/T001712/1) and Fondazione CARIPLO (2019-1836) for grant support.

# Competing interests

The authors report no competing interests.

# Ethics declaration

Written informed consent for genetic testing and publication of the relevant findings was obtained from all subjects. The study was conducted in accordance with the Declaration of Helsinki and was approved by the ethics committees in accordance with French ethics regulations [Paris Necker Ethics Committee approval (RBM 01-29 and RBM 03-48) to A.B. and A.D.].

# Supplementary material

Supplementary material is available at Brain online.

# Figure legends

**Figure 1. Flowchart for *RFC1* testing.**

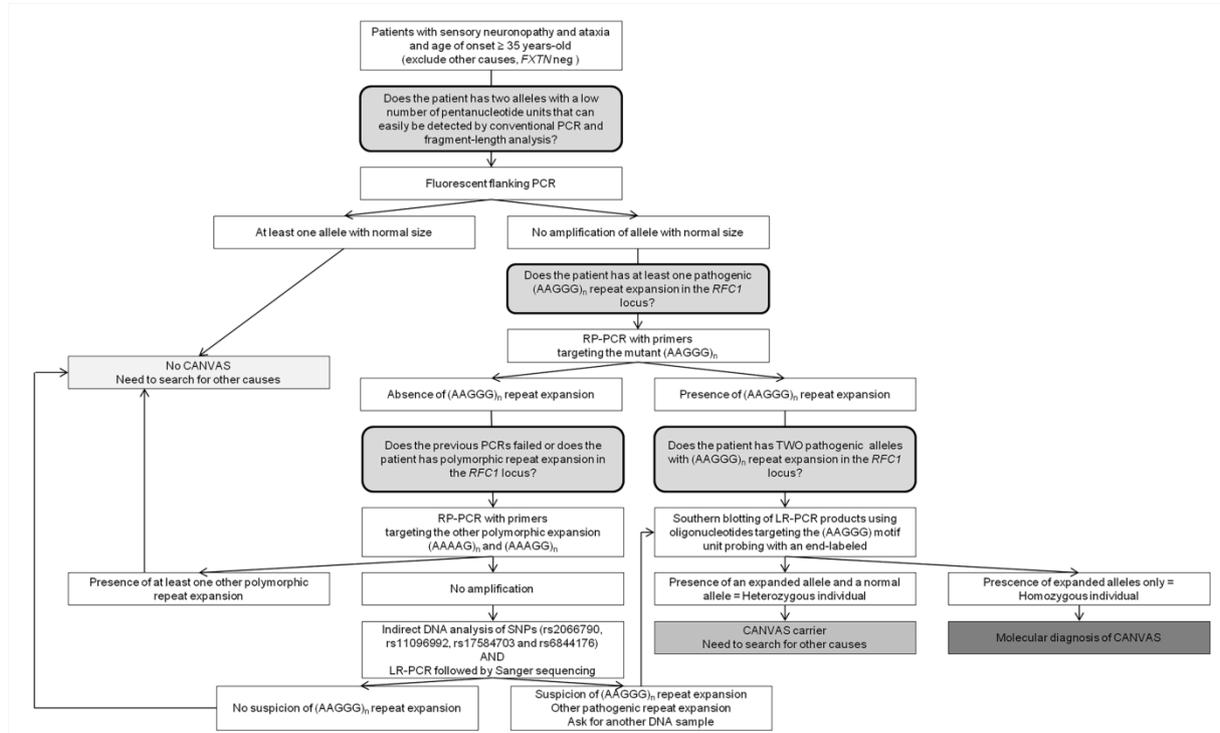

Workflow for *RFC1* expansion screening. Fluorescent flanking PCR is first used as a screening test. Then RP-PCR is performed to detect the $(AAGGG)_n$ repeat expansion. LR-PCR with Southern-blot revelation is performed to confirm the biallelic pathogenic expansion or detect other polymorphic expansions. Indirect analyses and Sanger sequencing is recommended in doubtful cases to limit the risk of false negatives linked to other pathogenic motifs.

**Figure 2. LR-PCR with Southern-blot revelation of different CANVAS patients or carriers.**

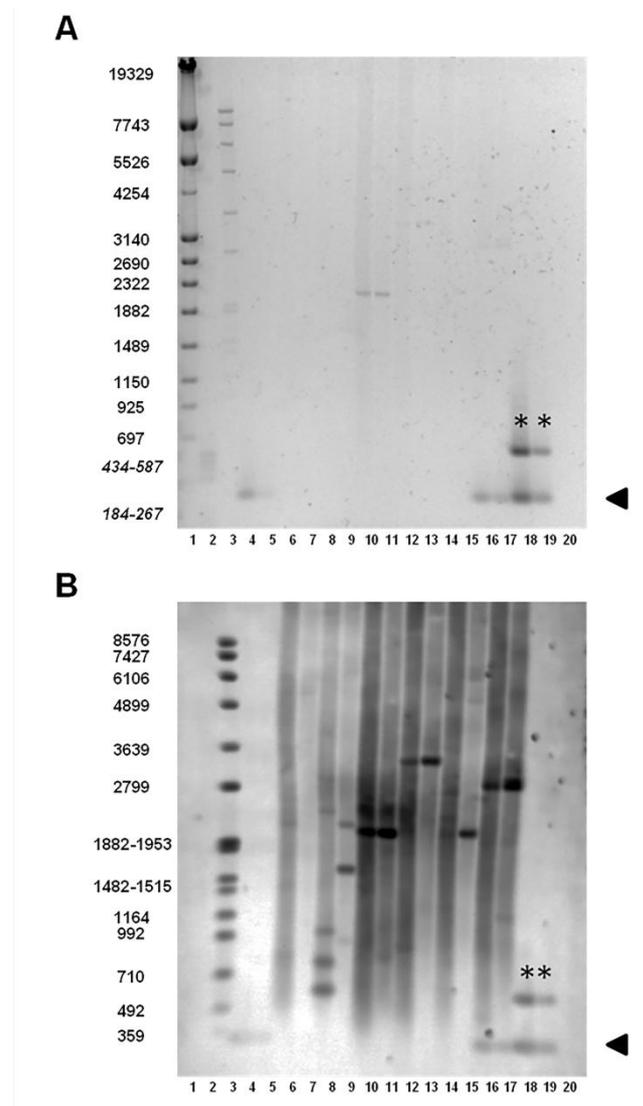

(**A**) Ethidium bromide-stained 1% agarose gel with DNA markers IV, V, and VII (lanes 1, 2, and 3, respectively). Lanes 4-5: healthy individuals with a normal allelic product band at 348 bp. Lanes 6-15: CANVAS patients. Lanes 16-17: Carrier. Lanes 18-19: healthy individuals with one polymorphic expansion. Lane 20: reagent blank. DNA for all individuals was loaded with two different inputs for LR-PCR (1 ng and 0.2 ng). The normal allelic product band at 348 bp is indicated by the arrow. The expanded polymorphic alleles are indicated by an *. (**B**) Southern blot of the gel shown above, probed with a $(CCCTT)_5$ probe. Only the digoxin-labelled Marker VII is visible in lane 3. The normal allelic product band at 348 bp is indicated by the arrow. The expanded polymorphic alleles are indicated by an *.

**Figure 3. Neuropathological examination of an RFC1 patient (KEN-334-6) carrying the homozygous AAGGG repeat expansion.**

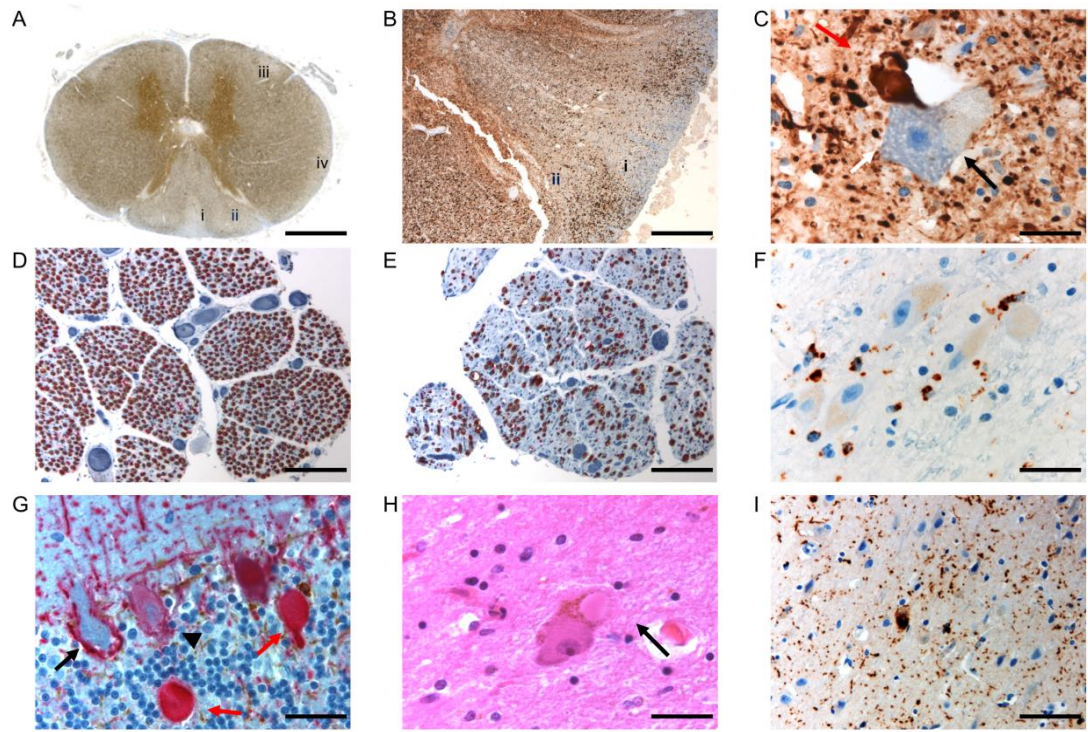

(**A**) Transversal section of the spinal cord at the upper thoracic level, phosphorylated neurofilament staining: alteration of the posterior columns, more prominent in the fasciculus gracilis (i) than the fasciculus cuneatus (ii), milder alteration of the vestibulospinal (iii) and spinocerebellar (iv) tracts. Scale bar = 2 mm. (**B**) A closer view of the posterior columns (SMI-310 IHC), showing greater pallor of the fasciculus gracilis (i) than the fasciculus cuneatus (ii) tract. Scale bar = 400 μm. (**C**) SMI-310 IHC showing axonal swelling (red arrow) at the contact with a motor neuron body (black arrow) in the anterior horn at the lumbar level. Scale bar = 20 μm. (**D-E**) MBP/2F11 staining showing normal anterior lumbar roots (**D**) and atrophic posterior lumbar roots (**E**). (**F**) CD68 IHC showing microglial activation in the vestibular nucleus. Scale bar = 80 μm. (**G**) Cerebellum, MBP/2F11 IHC preserved Purkinje cell (arrowhead), loss of a Purkinje cell shown by an empty basket (black arrow), and the presence of torpedoes (red arrows). Scale bar = 80 μm. (**H**) H&E staining showing a Lewy body in the *substantia nigra*. Scale bar = 80 μm. (**I**) α-synuclein staining showing the presence of diffuse α-synuclein in the entorhinal cortex. Scale bar = 40 μm.

**Figure 4. Astroglial abnormalities in an RFC1 patient (KEN-334-6) carrying the homozygous AAGGG repeat expansion.**

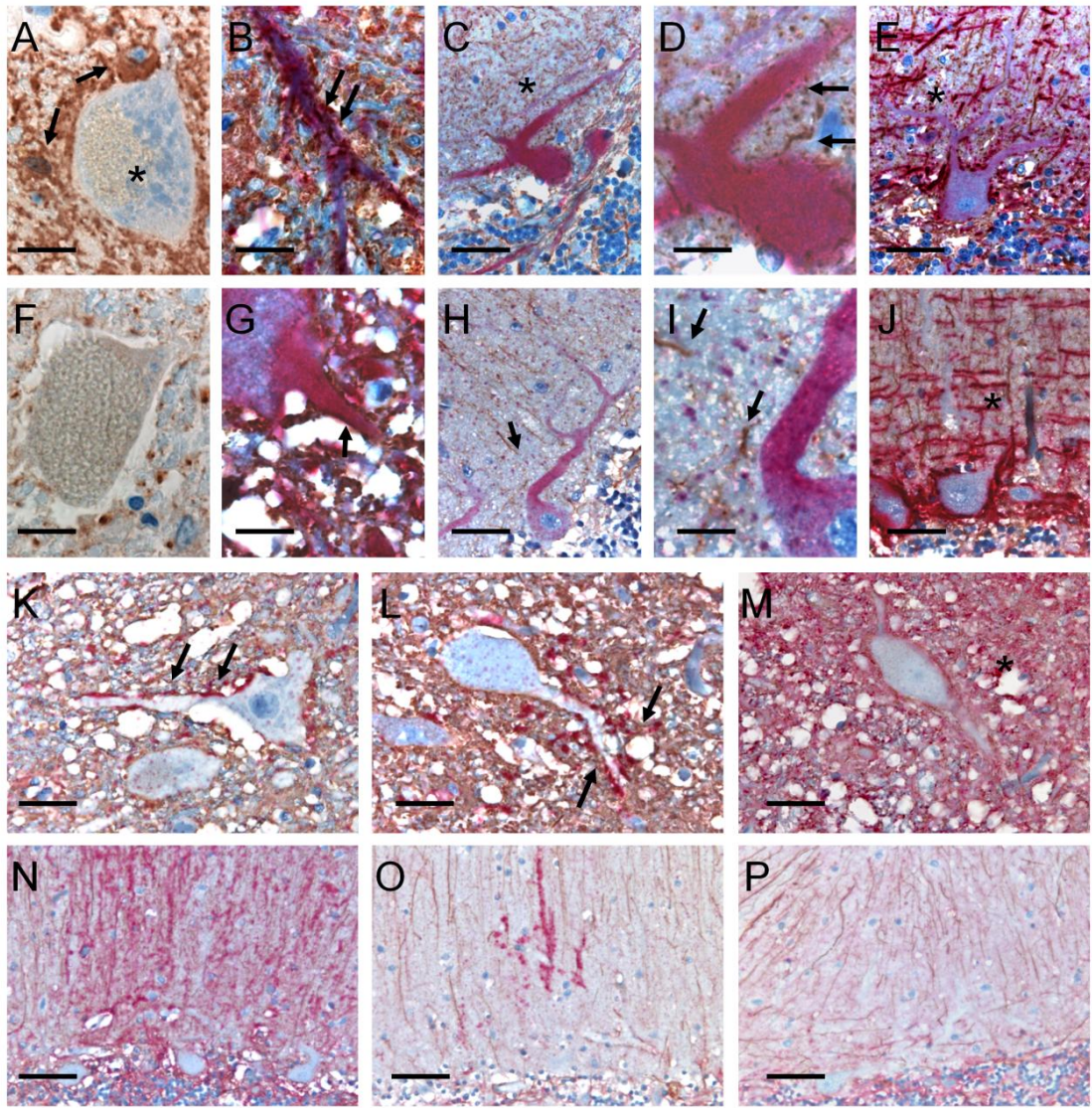

Neuron-astroglial pattern in the anterior horn of the spinal cord (A, B, F, G) and in the cerebellum (C-E, H-J) in the KEN-334-6 patient (A-E) and in a control without neurological disease (F-J). A and F: Immunohistochemistry of GFAP in the anterior horn (scale bars = 8 μm). In the patient, protoplasmic astrocytes (arrows) are in direct contact with the cell body of the apparently healthy motoneurons (*), which contrasts with the thin astrocytic framework in the control case (F). Double labeling of GFAP (brown) and non-phosphorylated neurofilaments (SMI-32, red) shows numerous astrocytic processes that appear to surround the patient's dendrites (B: arrows) compared to the control (G: arrow). Purkinje cell layer in the patient (C-E) compared to the control (H-J). Double labeling of GFAP (brown) and SMI-32 (red) shows in the patient a densification of GFAP immunoreactivity (C, D,*) and a loss of radial organization as compared to the control (H, I, arrows). As for the motor neurons, numerous astrocytic processes appear to surround the patient's dendrites (D, arrow). Double GFAP (brown) and phosphorylated neurofilaments (SMI-310, red) labeling shows a disorganization of parallel fibers in patient (E, *) compared to the control (J, *). Double GFAP (brown) and AQP4 (red) labeling in the anterior horn (K, L, M) and cerebellum (N, O, P) in the KEN-334-6 patient (K, N), an ALS control case (L, O) and a normal control (M, P). In the KEN patient AQP4 accumulates in astrocytic extensions in contact with motor neurons (K, arrows) similarly to what is observed in ALS (L, arrows). In the KEN-334-6 patient, as in ALS, AQP4 immunoreactivity is decreased in the anterior horn neuropile compared to the normal control case (M, *). In the molecular layer of the cerebellum, AQP4 immunostaining involves numerous blistered astrocytic processes (N), few in ALS (O) and absent in the control case (P). The age at post-mortem examination were 74, 71, and 95 in KEN-334-6 patient, ALS patient, and healthy control respectively. Scale bars = 2 μm (B, G); 8 μm (A, D, F, I), 20 μm (C, H, E, J-M) or 40 μm (N, O, P).

Table 1. Core CANVAS features and additional clinical signs of 38 homozygous carriers of pathogenic RFC1 AAGGG expansions (according to the disease duration of the index case).

| ID/sex | Age at onset (y) | Disease duration (y) | SARA (Max value 40) | Disability stage (Max value 7) | CANVAS features Cerebellar | Sensory | Vestibular | Chronic cough | First Motor neuron signs | Second motor neuron signs |
|---|---|---|---|---|---|---|---|---|---|---|
| AAD-1016-1/M* | 62 | 1 | 4 | 2 | CA | Vib, NP, SN | n/a | Yes | + | Fasciculations, myokymia |
| AAD-1016-2/W | (47), Syst exam | Syst exam | 4 | 1 | OM | Vib | n/a | n/a | + | Deficit, wasting |
| AAD-1016-4/W | 50 | 13 | 5 | 2 | OM | Vib, NP, SN | n/a | Yes | + | - |
| SAL-895/M* | 57 | 3 | n/a | 2 | Dys, OM, CA | Vib, SN | n/a | n/a | + Spasticity | - |
| CMT-1673/W* | 76 | 4 | n/a | 3 | CA | Vib, NP, SN | BVA | No | - | - |
| CMT-1550-1/W* | 44 | 4 | - | 0 | - | Vib, NP, SN | n/a | Yes | - | - |
| AAR-276/W* | 50 | 6 | n/a | 4 | OM | Vib, SN | n/a | n/a | + | - |
| PAD-253/M* | 64 | 6 | n/a | 3 | OM | Vib, SN | aVVOR, BVA | Yes | + | - |
| AAR-734/W* | 55 | 7 | 2 | 2 | OM | Vib, NP, SN | n/a | Yes | - | - |
| AAR-735-1/W* | 67 | 7 | 5 | 3 | Dys | Vib, NP, SN | n/a | Yes | - | - |
| AAR-642-4/M* | 45 | 8 | 10.5 | 3 | Dys, OM, CA | Vib, NP, SN | aVVOR | Yes | + Spasticity | Fasciculations, wasting, myokymia |
| AAR-642-5/W | 60 | 1 | 7 | 3 | Dys | Vib, SN | n/a | Yes | - | Cramps, myokymia |
| CMT1300/M* | 53 | 8 | 2 | 2 | OM | Vib, NP, SN | BVA | Yes | - | - |
| Z393/W* | 65 | 9 | 17 | 3 | CA | Vib, NP, SN | n/a | Yes | - | - |
| CMT-1537/W* | 46 | 10 | 3 | 3 | OM | Vib, NP, SN | n/a | Yes | + | - |
| CMT-1557-2/W* | 52 | 10 | 4 | 2 | OM | Vib, SN | n/a | Yes | - | Cramps, MN |
| CMT-1557-1/M | 60 | 0 | n/a | 1 | OM | Vib, SN | n/a | Yes | + | Cramps, fasciculations, MN |
| AAD-990/M* | 44 | 12 | 13 | 3 | OM, CA | Vib, SN | n/a | Yes | - | Cramps, fasciculations |
| A2730/M* | 50 | 12 | 6 | 3 | | Vib, SN | - | Yes | + | - |
| A3006/W* | 46 | 13 | 3 | 2 | OM | Vib, NP, SN | aVVOR, BVA | Yes | - | - |
| A2201/M* | 50 | 13 | 16 | 3 | Dys, OM, CA | Vib, SN | n/a | Yes | + | - |
| SAL-940/M* | 50 | 15 | 23.5 | 5 | Dys, OM, CA | Vib, NP, SN | n/a | n/a | - Spasticity | - |
| SAL-399-1021/M* | 69 | 16 | n/a | | OM, CA | Vib, SN | n/a | Yes | - | - |
| A1558/M* | 72 | 16 | 22 | 5 | Dys | Vib, SN | BVA | Yes | - | Fasciculations, wasting, MN |
| SAL-399-888/M* | 48 | 18 | n/a | 3 | Dys, CA | Vib, SN | n/a | Yes | + | - |
| SAL-1023/M* | 48 | 18 | 25 | 5 | Dys, OM, CA | Vib, SN | aVVOR | Yes | + | Cramps, myokimia |
| SAL-966/M* | 56 | 18 | 22 | 5 | Dys, OM, CA | Vib, NP, SN | n/a | Yes | + | Fasciculations, MN |
| A1977/M* | 59 | 19 | 26 | 5 | Dys, OM, CA | Vib, SN | BVA | Yes | + | - |
| AFT-098/M* | 40 | 20 | n/a | 4 | Dys, OM | Vib, NP, SN | aVVOR | Yes | + | - |
| AAR-680-1/W* | 48 | 23 | 22 | 5 | Dys, OM, CA | Vib, SN | n/a | Yes | - | - |
| AAR-520-10/W* | 50 | 23 | 22 | 5 | Dys, OM, CA | Vib, SN | n/a | Yes | + | - |
| AAD-1073-1/W* | 50 | 23 | 17 | 4 | Dys, OM, CA | Vib, NP, SN | BVA | Yes | - | Fasciculations, cramps, MN |
| AAD-417-6/M* | 50 | 28 | 10 | 4 | OM | Vib, NP, SN | n/a | Yes | - | Fasciculations |
| AAD-835-20/W* | 57 | 34 | 22 | 5 | Dys, OM, CA | Vib, NP, SN | n/a | Yes | - | MN |
| AAD-835-12/W | 67 | 13 | 27 | 5 | Dys, OM, CA | Vib, SN | n/a | Yes | - | MN |
| AAD-835-22/M | 58 | 20 | 9 | 3 | Dys, OM, CA | Vib, NP, SN | BVA | Yes | - | - |
| AAR-727/W* | 30 | 40 | 15 | 4 | Dys, OM, CA | Vib, NP, SN | aVVOR | Yes | - | Myokymia |
| KEN-334-6/M* | 25 | 49 | n/a | 5 | Dys, OM, CA | Vib, NP, SN | BVA | Yes | + | Fasciculations, cramps |
| n = 38 | 53.3±10.6 | 14.6±10.7 | 12.7±8.6 | 3.3±1.3 | | | | | | |

Abbreviations: +: present, -: absent, aVVOR: abnormal visually enhanced vestibulo-ocular reflex, BD: bladder dysfunction, BVA: bilateral vestibular areflexia, CA: cerebellar atrophy (brain MRI), Dys: cerebellar dysarthria, EPR: extensor plantar reflex, M: man, MN : motor neuropathy (nerve conduction studies), n/a: not available, NP: neuropathic pain, OM: oculomotor abnormalities, SN: sensory neuronopathy (nerve conduction studies), Vib: decreased vibration sense at ankles, W: woman, y: years.

* Proband in each family.

**Table 2. Main differential diagnoses of CANVAS.**

| Disease | CANVAS (Our cohort) | Very late onset Friedreich Ataxia (Lecocq 2016)[22] | POLG-related disorders (Tchikviladzé 2015)[34] |
|---|---|---|---|
| Gene | *RFC1* biallelic AAGGG expansion | *FXN* < 500 GAA biallelic expansion | *POLG*[+/+] |
| Transmission | Recessive | Recessive | Recessive |
| Age at onset (years) | 54 (30–76) | 49 (40–78) | 31 (2–61) |
| Sensory Neuropathy | 97% | 47% | 63% |
| Decreased/abolished reflexes | 75.6% at ankles 27% at patella 19% at upper limbs | 20% | 85% |
| Cerebellar dysarthria | 51% | 3% | 35% |
| Vestibular areflexia | 86%[a] | Usually absent | Usually absent |
| First motor neuron involvement | 43%[b] | 55%[b] | 11% |
| Dysautonomia | 48%[a] | Usually absent | Usually absent |
| Cerebellar atrophy | 72%[a] | Usually absent | 35% |
| Other neurological and extraneurological features | Cough 91%[a] Skeletal deformities 21% Movement disorders 18% Ptosis 8% | Cardiomyopathy 11% Foot deformities 10% | Ophthalmoplegia 83% Ptosis 80% Movement disorders 53% Psychiatric symptoms 41% Hypoacusia 39% Cognitive impairment 32% Epilepsy 17% Optic atrophy, retinitis pigmentosa, cataract Gastrointestinal pseudo-obstruction |

[a] These three late-onset diseases present sensory neuronopathy and cerebellar ataxia as the main clinical features. However, certain features make it possible to distinguish CANVAS from very late onset Friedreich ataxia or POLG-related disorders, such as vestibular areflexia, chronic cough, dysautonomia, and cerebellar atrophy.

[b] First motor neuron involvement is similar between CANVAS and very late onset Friedreich ataxia.

# Supplementary material

## Supplementary methods

### Targeted sequencing (Ataxia panel)

Exon capture of the 359 selected genes and libraries was performed using the Haloplex kit (Agilent Technologies) according to the manufacturer's instructions. Pooled libraries ($n = 8$) prepared using the ataxia panel were sequenced on a MiSeq system (Illumina, San Diego, CA, USA) with MiSeq Reageant kit v2.

Sequence analysis was performed using SeqPilot (v4.3.1, JSI Medical Systems, Ettenheim, Germany), MiSeqReporter software (v2.6, Illumina, San Diego, CA, USA), and Genesearch (PhenoSystems, Braine le Chateau, Belgium) software. Variants were annotated and filtered using Variant Studio software (v3.0, Illumina, San Diego, CA, USA). Harmful effects were predicted using Alamut software (v2.9.0, Interactive BioSoftware, Rouen, France).

### Exomes

#### Exome sequencing

Samples were prepared using a SeqCap EZ Human Exome Probes v3.0 capture kit (Roche) and sequenced using paired-end, 150-cycle chemistry on an Illumina NextSeq500 (Illumina, San Diego, CA) according to the manufacturer's protocol.

#### Bioinformatics analysis

The sequence data were aligned to the hg19 assembly version of the human genome using BWA v0.7.17.[37] Variant calling, joint genotyping, and recalibration were realized performed using GATK v3.7.[38] Variant annotation was performed using Ensembl Variant Effect Predictor (VEP) (version 101, Assembly: GRCh37.p13)[39], Only point variants and small indels (< 50 bp) were investigated with this pipeline.

#### Variant filtering

For the search of candidate genes for the seven patients without CANVAS and putative modifier variants in the 36 CANVAS patients, we used the following filters: (i) quality filter (PASS GATK filter, variant coverage ≥10), (ii) heterozygosity (allelic frequency between 0.3 and 0.7), (iii) effect on coding sequence (exonic non-synonymous or canonical splice variant), (iv) frequency in the internal database (allele count ≤ 2 amongst the 250 exomes from subjects without neurological disorders; *ie*, ≤ 0.8%), (v) frequency in the public databases: for variants compatible with recessive inheritance (homozygous variants or double heterozygous variants in the same gene), we considered only those with an allele count < 3 at the homozygous/hemizygous state in the gnomAD database [http://gnomad.broadinstitute.org/]; for variants compatible with dominant inheritance, we considered only those with an allele count < 5, according to VEP.

Candidate genes were prioritized on the basis of potentially damaging variants and the effect on the protein coding transcript (exclusion of immunoglobulin transcripts, non-coding transcripts, nonsense-mediated decay transcripts, and pseudogenes) using prediction tools from VEP and the dbNSFP database[40], including Condel, CADD, DEOGEN2, Eigen-PC-phred_coding, FATHMM, LRT, M-CAP, MetaLR, MetaSVM, MutationAssessor, MutationTaster, PROVEAN, Polyphen2, PrimateAI, SIFT, Fathmm-MKL, and phastCons100way_vertebrate. Rare variants in uncommon alternative transcripts and in-frame deletion/insertions could not be tested by certain prediction tools. Thus, we manually curated all variants with more than four missing predictions. We analyzed rare variants in all disease-causing genes in the recent literature and in OMIM, including a neurological phenotype with adult onset. Finally, we classified the variants according to the ACMG classification of the genetic variants[41] using Varsome (https://varsome.com/).[42] We kept only the variants classed as "Uncertain significance", "Likely pathogenic", or "Pathogenic".

For the seven patients without CANVAS, we also added the following filters. (vi) Variants were filtered when their corresponding gene was not expressed in the brain or spinal cord in GteEX (https://www.gtexportal.org/home/) and Allan Brain Atlas (https://www.brain-map.org/) databases. (vii) We excluded the variants in genes belonging to the top 100 most frequently mutated genes (FLAGS) in public exomes according to Shyr *et al*.[43] (viii) For loss-of-function variants, we considered only variants in genes in which loss-of-function variants were a previously reported mechanism of disease or with a probability of having a loss-of-function intolerance score > 0.9 and an upper bound of confidence interval < 0.35 for the observed / expected loss-of-function variants metric in the gnomAD database.[44] (ix) When available, segregation of the variants was assessed in family members. All relevant variants and variants of unknown significance were verified by Sanger sequencing.

### Indirect *RFC1* analysis by SNP genotyping

In the case of absence of amplification using different RP-PCRs, we propose to genotype the SNPs rs2066790, rs11096992, rs17584703, and rs6844176 reported by Cortese *et al.*[1], for indirect *RFC1* analysis. This analysis is useful to exclude the classical recessive haplotype shared by patients with the classical pathogenic expansion (AAGGG)$_n$ (rs2066790-AA, rs11096992-AA, rs17584703-GG, and rs6844176-AA). SNPs are genotyped using PCR followed by Sanger sequencing. We used the same primers concentrations, and PCR thermocycling conditions described previously for all SNPs.[1] Primers sequences are:

rs2066790-Fw = 3'-CCTGAGGTGTGTGGCTTTAG-5'

rs2066790-Rv = 3'-TCAGGACTTACAGACTTTGGGA-5'

rs11096992-Fw = 3'-TGGCTTAAATGATCTTTTCCCG-5'

rs11096992-Rv = 3'-CACCAATAAAACTTACACCCACA-5'

rs17584703-Fw = 3'-CCCTTTCGAAATTTGAACCACG-5'

rs17584703-Rv = 3'-GGAGTGGAGCAATGAAACAGTT-5'

rs6844176-Fw = 3'-ACCCACATCGATGCAGTTTTAC-5'

rs6844176-Rv = 3'-TGCCCAAGACCACGTAACTATT-5'

### Sanger sequencing of long-range PCR products

After long-range PCR amplification as described in materials and methods, we performed Sanger sequencing of PCR products similarly to Cortese *et al.*[1] using the same primers for sequencing.

## Supplementary data

### Additional variants in *RFC1* biallelic expansion carriers

Previous molecular screening using targeted NGS or exome sequencing was inconclusive in our cohort, but revealed additional rare variants in 12 OMIM genes that may influence the phenotype. According to the ACMG classification, three variants in genes were pathogenic, one missense previously reported in *POLG* NM_001126131.2:c.925C>T, p.(Arg309Cys), a nonsense variant in *TWNK* gene: NM_021830.5:c.624G>A, p.(Trp208*), and one frameshift variant in *RAPGEF2*: NM_001351724.1:c.4194_4195insGCCCCCCC p.(Pro1399Alafs*68). The others variants were of uncertain significance with moderate impact on protein function (Supplemental Table S5).

## Variants in non-RFC1 expanded patients

One patient (AAD-604-1) carried a heterozygous *RFC1* expansion and a missense variant, c.2663C>G p.(Ser888Cys), in *GRM1*, causing spinocerebellar ataxia type 44 (SCA44, OMIM #617691) (Supplemental Table S3). The phenotype associated ganglionopathy with optic neuropathy, early cataracts, ophthalmoplegia, dysphagia, parkinsonism, *pes cavus*, and bladder dysfunction. Until now, SCA44 has been reported as adult-onset cerebellar ataxia or early onset ataxia with intellectual disability in the presence of loss-of-function variants.[45]

Five patients had no expansions in *RFC1* (Supplemental Table S3). There were too many rare variants in these patients to define a new candidate gene. Only one rare variant with unknown significance in *MYH14* (NM_001145809.2:c.2003G>A (p.Gly668Glu)) could partially explain the phenotype, with an early hearing loss phenotype in the corresponding case. The rare variants found in non-RFC1 patients are listed in supplemental Table S4.

**Supplemental Figure S1. LR-PCR with Southern-blot revelation of patient KEN-334-6**

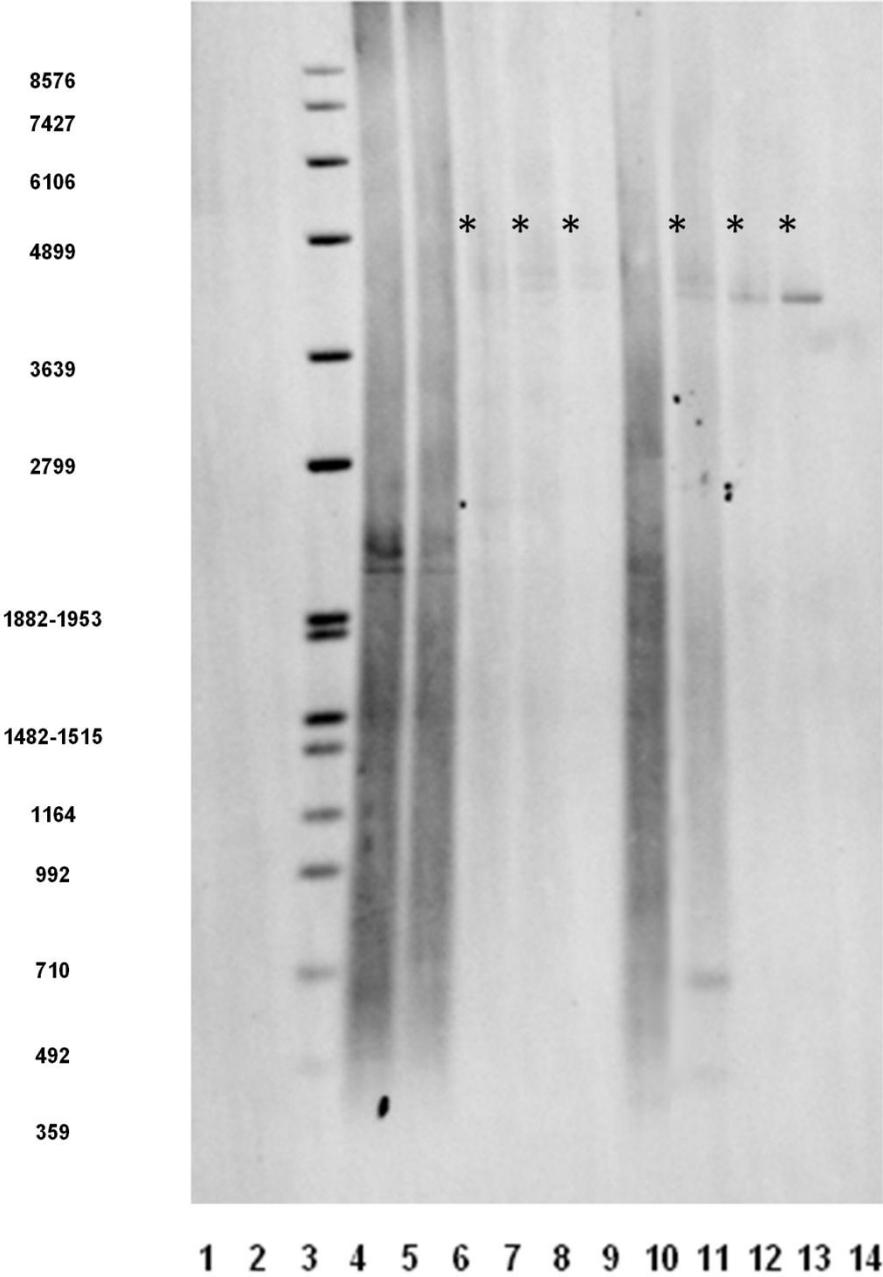

Southern blotting of long-range PCR products. DNA markers IV, V, and VII (lanes 1, 2, and 3, respectively, only the digoxin-labelled Marker VII is visible in lane 3). Lanes 4-8: Patient KEN-334-6, blood sample with different DNA input used for LR-PCR (from left to right: 10, 5, 2, 1, 0.2 ng). Lanes 9-12: KEN-334-6, cerebellum sample with different DNA input (from left to right: 5, 2, 1, 0.2 ng). Lane 13: reagent blank. Expanded alleles located approximately between 4050 to 4200 bp for the blood (≈750 to 790 repeats) and 4000 to 4050 bp for the cerebellum (≈740 to 750 repeats) are indicated by *.

**Supplemental table S1. PCR thermocycling conditions for LR-PCR**

| Step | Temperature | Time | Cycles |
| --- | --- | --- | --- |
| Initial denaturation | 94°C | 1 min | 1 |
| Denaturation | 94°C | 15 s | |
| Annealing | 59°C | 30 s | 10 |
| Extension | 68°C | 3 min | |
| Denaturation | 94°C | 15 s | |
| Annealing | 59°C | 30 s | 25 |
| Extension | 68°C | 3 min + 15 s/cycle | |
| Final extension | 68°C | 10 min | 1 |

**Supplemental table S2. Pretreatment and antibodies used for immunohistochemistry analyses.**

| Antigen | Poly/monoclonal | Producer | Immunogen | Clone | Pretreatment | Dilution | Incubation time |
|---|---|---|---|---|---|---|---|
| Aβ | Monoclonal (mouse) | Dako® | Deposits of beta-amyloid | 6F/3D | Formic acid | 1:200 | 120 min |
| Alpha-synuclein | Monoclonal (mouse) | Millipore® | N/A | 5G4 | CC1® | 1:4000 | 120 min |
| AQP4 | Polyclonal (rabbit) | Sigma® | Aquaporin 4 water channels in astrocytes membranes | | CC2® | 1:3000 | 32 min |
| CD163 | Monoclonal (mouse) | Cell Marque® | Acute phase-regulated transmembrane protein on monocytes | MRQ-26 | CC1® | 1:50 | 32 min |
| CD20 | Monoclonal (mouse) | Dako® | Transmembrane protein expressed on B cells | L26 | CC1® | 1:100 | 32 min |
| CD3 | Monoclonal (rabbit) | Ventana® | Epsilon chain of the human CD3 | 2GV6 | CC1® 30 min at 95°C | prediluted | 32 min |
| CD68 | Monoclonal (mouse) | Dako® | Lysosomal fraction of human macrophages | KP1 | CC1® | 1:100 | 20 min |
| GFAP | Monoclonal (mouse) | Dako® | Glial Fibrillary Acidic Protein | 6F2 | CC1® | 1:500 | 28 min |
| MBP | Monoclonal (rabbit) | BioSB® | Myelin basic Protein | EP207 | CC1® | 1:50 | 60 min |
| Neurofilament | Monoclonal (mouse) | Dako® | 70 kDa subunit of neurofilament (axons) | 2F11 | CC1® | 1:2000 | 32 min +37°C |
| Phosphorylated neurofilaments | Monoclonal (mouse) | Biolegend® | Phosphorylated neurofilaments H/M in axons | SMI-310 | CC1® | 1:4000 | 60 min |
| Non-phosphorylated neurofilaments | Monoclonal (mouse) | Biolegend® | Neurofilament heavy polypeptide, NF-H in neuronal cell bodies, and dendrites | SMI-32 | CC2® | 1:1000 | 60 min |
| p62 | Monoclonal (mouse) | BD Transduction laboratories | Human p62 lck ligand aa. 257-437 | 3/P62 LCK LIGAND | CC2® | 1:500 | 32 min |
| Tau | Monoclonal (mouse) | ThermoFisher® | Phospho-Tau (Ser202, Thr205) | AT8 | CC1® | 1:500 | 32 min |
| TDP43 | Rabbit polyclonal | Proteintech® | TAR DNA binding protein | _ | CC1® | 1:2000 | 80 min |

Abbreviations: ***CC1***, a proprietary high pH (=8) buffer; ***CC2***, pH (=6); ***FA***: formic acid.

**Supplemental Table S3. Clinical characteristics of non-*RFC1* patients.**

| ID/sex | Genetic status | ACMG | Age at onset (y) | Disease duration (y) | SARA (Max value 40) | Disability stage (Max value 7) | CANVAS features Cerebellar | CANVAS features Sensory | CANVAS features Vestibular | Chronic cough | First motor neuron | Second motor neuron signs | Plus phenotype |
|---|---|---|---|---|---|---|---|---|---|---|---|---|---|
| AAR-604-1/W* | *GRM1* c.2663C>G *RFC1* ht | US | 40 | 22 | 21.5 | 4 | Dys, OM, CA | Vib, NP, SN | n/a | Yes | - | Cramps | Optic neuropathy, early cataracts, ophthalmople, dysphagia, parkinsonism, cavus, BD |
| AAD-1030-1/M* | - | | 66 | 9 | 11 | 4 | Dys, OM, CA | Vib | | Yes | - | Deficit | Pes cavus, BD, hearing l, white matter abnormali |
| AAD-1030-2/M | - | | 64 | 6 | n/a | n/a | Dys, OM | Vib, NP, SN | n/a | No | - | Deficit | - |
| AAR-461-7/M* | *MYH14* c.2003G>A | US | 23 | 17 | 24 | 6 | Dys, OM, CA | Vib, SN | n/a | No | - | - | Early cataracts, optic ne excavation, hearing los camptocormia, scolios |
| AAR-508-13/W* | - | | 10 | 24 | 9 | 2 | Dys, OM, CA | Vib, SN | n/a | No | - | - | Myoclonic tremor, hearing dysphagia |
| SAL-399-926/M* | - | | 53 | 3 | 20.5 | 5 | Dys, OM, CA | Vib, SN | n/a | No | - | - | Parkinsonism, BD, dysph brainstem atrophy, cogn impairment |
| *n* = 6 | | | 42.7±22.7 | 13.5±8.7 | 14.6±8.1 | 4.2±1.5 | | | | | | | |

Abbreviations: absent, BD: bladder dysfunction, BVA: bilateral vestibular areflexia, CA: cerebellar atrophy (brain MRI), Dys: cerebellar dysarthria, EPR: extensor plantar reflex, ht: heterozygous, M: man, n/a: not available, NP: neuropathic pain, OM: oculomotor abnormalities, SN: sensory neuronopathy (nerve conduction studies), US: uncertain significance, Vib: decreased vibration sense at ankles, W: woman, y: years.

*Proband in each family.

**Supplemental table S4: List of candidate variants in non RFC1 patients.**

The first tab describes the different filters used. Each tab represents a filter. The last "output" tab lists all the additional variants. Abbreviations: HET: heterozygous, HET COMP: putative compound heterozygous, HEM: hemizygous, AD: autosomal dominant, AR: autosomal recessive, XLR, X-linked recessive, US: uncertain significance, LP: likely pathogenic, P: pathogenic. na: not applicable. All the abbreviations for the results of the different prediction tools are available at https://drive.google.com/file/d/1VINShZQYN63HvigJA_SiuOYIk8TWl5Fh/view

**Supplemental Table S5: List of variants found among *RFC1* pathogenic AAGGG expansions carriers.**

| ID/sex | Gene | Variant | Status | Inheritance of OMIM disorder | Impact | Pathogenicity predictions | ACMG Classification | Comments | Additional features |
|---|---|---|---|---|---|---|---|---|---|
| AAD-1016-1/M | *POLG* | NM_001126131.2:c.925C>T, p.(Arg309Cys) | HET | AD, AR | moderate | 16/17 | P | previously reported mutation | Peripheral motor neuron signs, parkinsonism |
| AAD-1016-2/W | *POLG* | NM_001126131.2:c.925C>T, p.(Arg309Cys) | HET | AD, AR | moderate | 16/17 | P | previously reported mutation | Peripheral motor neuron signs, epilepsy |
| AAD-1016-4/W | *POLG* | NM_001126131.2:c.925C>T, p.(Arg309Cys) | HET | AD, AR | moderate | 16/17 | P | previously reported mutation | |
| SAL-940/M | *TWNK* | NM_021830.5:c.624G>A, p.(Trp208*) | HET | AD, AR | high | 6/7 | P | | Postural tremor, parkinsonism, epilepsy |
| SAL-888/M | *RAPGEF2* | NM_014247.3: c.4194_4195insGCCCCCCC p.(Pro1399Alafs*68) | HET | AD | high | n.a | P | | Cervical spinal cord atrophy |
| SAL-888/M | *OPA1* | NM_130837.3:c.2648T>C, p.(Val883Ala) | HET COMP* | AD, AR | moderate | 17/17 | US | | |
| SAL-888/M | *OPA1* | NM_130837.3:c.2983+112T>C, p.? | HET COMP* | AD, AR | moderate | 2/13 | US | | Ptosis, cervical spinal cord atrophy |
| AFT-098/M | *SPTLC1* | NM_006415.4:c.388G>A, p.(Val130Met) | HET | AD | moderate | 15/17 | US | | |
| SAL-1021/M | *UBTF* | NM_014233.4:c.782G>A, p.(Arg261Lys) | HET | AD | moderate | 14/17 | US | | Cortical atrophy, action tremor upper limbs |
| AAD-417-6/M | *VARS2* | NM_001167734.1:c.1313A>G, p.(His438Arg) | HET COMP* | AR | moderate | 12/17 | US | | |
| AAD-417-6/M | *VARS2* | NM_001167734.1:c.1674G>T, p.(Trp558Cys) | HET COMP* | AR | moderate | 13/17 | US | | Peripheral motor neuron signs, hearing loss |
| AAD-417-6/M | *ADCY5* | NM_183357.3:c.202C>T, p.(Arg68Cys) | HET | AD | moderate | 9/17 | US | | Peripheral motor neuron signs |
| AAR-727/W | *COA6* | NM_001012985.2:c.289T>C, p.(Tyr97His) | HET COMP* | AR | moderate | 11/17 | US | | |
| AAR-727/W | *COA6* | NM_001206641.3:c.29C>T, p.(Pro10Leu) | HET COMP* | AR | moderate | 4/11 | US | | Peripheral motor neuron signs, parkinsonism, epilepsy |
| AAD-1016-1/M | *ATP7A* | NM_000052.7:c.548A>G, p.(His183Arg) | HEM | XLR | moderate | 11/15 | US | | Peripheral motor neuron signs |
| CMT-1537/W | *FIG4* | NM_014845.6:c.1120_1122del, p.(Ile374del) | HET | AD, AR | moderate | n.a | US | | Peripheral motor neuron signs, pes cavus, scoliosis |
| CMT-1557-1/M | *PRX* | NM_181882.3:c.1913_1915del, p.(Glu638del) | HET | AD, AR | moderate | n.a | US | | Peripheral motor neuron signs |